# Coherent laser spectroscopy of highly charged ions using quantum logic

P. Micke[1,2*], T. Leopold[1*], S. A. King[1*], E. Benkler[1], L. J. Spieß[1], L. Schmöger[1,2], M. Schwarz[1,2], J. R. Crespo López-Urrutia[2], P. O. Schmidt[1,3]

[1] Physikalisch-Technische Bundesanstalt, Bundesallee 100, 38116 Braunschweig, Germany.

[2] Max-Planck-Institut für Kernphysik, Saupfercheckweg 1, 69117 Heidelberg, Germany.

[3] Institut für Quantenoptik, Leibniz Universität Hannover, Welfengarten 1, 30167 Hannover, Germany.

*These authors contributed equally to the work.

**Precision spectroscopy of atomic systems[1] is an invaluable tool for the advancement of our understanding of fundamental interactions and symmetries[2]. Recently, highly charged ions (HCI) have been proposed for sensitive tests of physics beyond the Standard Model[2–5] and as candidates for high-accuracy atomic clocks[3,5]. However, the implementation of these ideas has been hindered by the parts-per-million level spectroscopic accuracies achieved to date[6–8]. Here, we cool a trapped HCI to the lowest reported temperatures, and introduce coherent laser spectroscopy on HCI with an eight orders of magnitude leap in precision. We probe the forbidden optical transition in $^{40}\text{Ar}^{13+}$ at 441 nm using quantum-logic spectroscopy[9,10] and measure both its excited-state lifetime and *g*-factor. Our work ultimately unlocks the potential of HCI, a large, ubiquitous atomic class, for quantum information processing, novel frequency standards, and highly sensitive tests of fundamental physics, such as searching for dark matter candidates[11] or violations of fundamental symmetries[2].**

Alike a microscope aimed at the quantum world, laser spectroscopy pursues ever higher resolving power. Every increase in resolution enables deeper insights into the subtle effects that all known fundamental interactions have on the atomic wave function. Advances in optical frequency metrology have dramatically improved resolution in the last three decades[1], and are making laser spectroscopy an extremely sensitive tool for studying open physics questions such as the nature of dark matter, the strength of parity violation, or a possible violation of Einstein's theory of relativity[2]. However, only a few atomic and ionic species are currently within the reach of cutting-edge optical frequency metrology. Expanding this field of exploration to systems with high sensitivity to such effects is therefore crucial. Due to their extreme properties, highly charged ions (HCI) are promising candidates for such fundamental tests. Contributions from special relativity, quantum electrodynamics (QED), and the nucleus to the binding energies of their outer electrons become several orders of magnitude larger than in neutral atoms. This renders them ideal systems for benchmarking the most advanced theories and calculations, which has been repeatedly demonstrated via optical fluorescence spectroscopy in electron beam ion traps (EBIT)[6,7], x-ray spectroscopy in storage rings[12] and EBITs[13–15], and ground-state *g*-factor studies in Penning traps[16,17]. The hyperfine spitting of the 1s state in heavy hydrogen-like ions can even shift into the optical range, providing laser-accessible transitions (see, e.g., [18–20]) with nuclear size contributions on the order of several percent of the total transition energy.

It was realised recently that a non-gravitational coupling of dark matter to ordinary matter would affect atomic energy levels[11], and thus become observable in optical clock comparisons as an apparent drift or modulation of the fine-structure constant $\alpha$. HCI offer narrow-linewidth optical transitions that are among the most sensitive to a possible variation of $\alpha$[4]. In addition, their inherent insensitivity to external electric fields[3] leads to significantly smaller systematic perturbations compared to neutral and singly charged atoms. This makes them potentially superior references for high-accuracy optical atomic clocks, with many proposed species reviewed in reference [5]. However, so far, no experiment has been capable of performing laser spectroscopy at the required level of precision. The major limitation was set by the high temperature of a few million kelvins at which HCI are produced and typically stored. This induces Doppler broadenings with full-width-at-half-maximum (FWHM) linewidths of several tens of gigahertz, and corresponding line-centre uncertainties of a few hundreds of megahertz in the best cases[6–8]. Since HCI generally do not offer suitable transitions for direct laser cooling, sympathetic cooling of multiple HCI by laser-cooled $^9Be^+$ ions was implemented in a Penning trap at the Lawrence Livermore National Laboratory[21], reaching an ion temperature of around 4 K. More recently, the Cryogenic Paul Trap Experiment (CryPTEx)[22] demonstrated reliable Coulomb crystallization of single $^{40}Ar^{13+}$ ions in a crystal of many $^9Be^+$ ions. Sympathetic Doppler cooling down to the 10 mK level and two-ion crystal preparation[23,24] paved the way for high-accuracy spectroscopy. Even so, spectroscopy of narrow transitions in single ions requires efficient state detection on a different, fast cycling transition[1], typically also used for laser cooling. If such a transition is not available, quantum logic spectroscopy (QLS) can be employed[9,10]. There, the 'spectroscopy ion' (in this case the HCI) is co-trapped with a so-called 'logic ion' ($^9Be^+$) that provides sympathetic cooling, state preparation, and is used for state detection. These functions are enabled by means of the Coulomb interaction between both ions, allowing lasers to couple their internal electronic levels with their quantised joint motion in the trap. This technique and variations thereof have been successfully employed for optical atomic clocks based on $Al^+$ ions[25–27], for internal state detection and spectroscopy of molecular ions[28,29], and for spectroscopy of broad transitions in atomic ions[30,31].

Here, we demonstrate for the first time QLS of an HCI, specifically of the electric dipole-forbidden transition between the $^2P_{1/2}$ and $^2P_{3/2}$ fine-structure levels of $^{40}Ar^{13+}$ at a wavelength of 441 nm, the most accurately known transition in any HCI. We achieve a FWHM well below 100 Hz, close to the natural linewidth of 17 Hz. Single line scans taken on a time scale of a few minutes determine the line centre with an uncertainty below 2 Hz. This corresponds to a fractional statistical uncertainty of $3\times10^{-15}$ for the transition frequency of approximately 680 THz and compares favourably to previous measurements taken over hours or even days to achieve $2\times10^{-7}$ relative uncertainties[6–8]. Quantum logic-assisted state preparation of the $^{40}Ar^{13+}$ ion allows us to measure all six Zeeman components of the transition, which split up on a megahertz scale in the 160 µT magnetic quantisation field. This allows us to determine the $g$-factor of the $^2P_{3/2}$ excited state with unprecedented accuracy. Furthermore, we demonstrate a quantum logic-assisted excited-state lifetime measurement.

## Preparation of a single highly charged ion

A detailed description of the experimental setup is given in Methods (see also Extended Data Figs. 1 and 2). In brief, argon HCI are produced by an electron beam ion trap (EBIT), PTB-EBIT[32], and ejected from it in triggered bunches of ~200 ns duration with a mean kinetic energy of approximately 700 $q$V, where $q$ is the ion charge. The HCI are guided to the spectroscopy trap through an ion-optical beamline. Based on their time-of-flight, we select the $^{40}Ar^{13+}$ ions by rapidly switching a gate electrode. A pulsed gradient potential decelerates them electrodynamically[33] to about 146 $q$V. Then, a single $^{40}Ar^{13+}$ ion stochastically enters the cryogenic linear Paul trap[34]. The trap is globally biased to +138 V, thereby slowing the HCI down to 8 $q$V upon entry. After passing through the trapping region, the HCI is reflected back by an electrode at the end of the Paul trap. A mirror electrode in front of the trap is switched up to prevent the ion from escaping again, thereby capturing the HCI in an oscillatory axial motion. The repeated crossing through a pre-prepared laser-cooled $^9Be^+$ Coulomb crystal within the trap dissipates the HCI's residual kinetic energy. After sufficient sympathetic cooling, the $^{40}$Ar$^{13+}$ ion joins the Coulomb crystal. Excess $^9$Be$^+$ ions are deliberately removed until a two-ion crystal has been prepared (see Fig. 1). The entire two-ion crystal preparation procedure takes only a few minutes. The Paul trap is refrigerated to below 5 K by a mechanically decoupled, closed-cycle cryostat to provide a vacuum below the 10$^{-12}$ Pa level (corresponding to a particle density < 20000/cm$^3$), thus suppressing charge-exchange collisions and achieving HCI storage times on the order of 45 minutes[35].

## Ground-state cooling and quantum logic

The implementation of QLS requires control and preparation of the motional and internal states of both ions using coherent laser pulses on carrier and sideband transitions. After the two-ion crystal preparation, the strong Coulomb coupling between the two ions results in joint motional modes within the trap. Sympathetic cooling, state preparation and QLS are performed by repeating the experimental sequence shown in Fig. 2. First, Doppler cooling and optical pumping on the $^9$Be$^+$ $^2$S$_{1/2}$-$^2$P$_{3/2}$ cycling transition (see Fig. 2a) are applied. The two axial normal modes of the Coulomb crystal with secular frequencies of about $\nu_{IP}$ = 1.37 MHz (in-phase) and $\nu_{OP}$ = 1.86 MHz (out-of-phase) respectively, are then cooled to the quantum mechanical ground state of motion with final average occupation numbers of $\bar{n}$ = 0.05 (IP) and 0.02 (OP), corresponding to an effective temperature of less than 50 µK for each mode. For this purpose, we use laser pulses which coherently couple the electronic degrees of freedom to the common motional modes (a technique referred to as resolved sideband cooling). To do this, stimulated Raman transitions are driven between the $^9$Be$^+$ hyperfine qubit states $|\downarrow\rangle_L|n\rangle_m \rightarrow |\uparrow\rangle_L|n-1\rangle_m$ (where $|n\rangle_m$ denotes the motional quantum state of the IP or OP modes) using two laser beams with a wavelength of 313 nm. In our low magnetic field, the states are separated by a frequency of approximately 1.25 GHz. A repumping laser couples the $^9$Be$^+$ $^2$S$_{1/2}$ and $^2$P$_{1/2}$ levels (not shown in Fig. 2a) for electronic state preparation and for depopulation of state $|\uparrow\rangle_L$ [36]. The Ar$^{13+}$ Zeeman ground state is then deterministically prepared with clock laser sideband pulses (see Extended Data Fig. 3). After full state preparation, QLS[9] is performed in four steps (see Fig. 2): A clock laser pulse of tuneable length and power is applied (1), which couples the ground and excited state in Ar$^{13+}$ coherently. After a variable wait time for excited-state lifetime measurements, a clock laser red sideband π-pulse maps the excitation from the electronic Ar$^{13+}$ state onto the common axial OP mode (2) and another red sideband π-pulse on the $^9$Be$^+$ hyperfine qubit transition maps it onto the $^9$Be$^+$ electronic state (3). Finally, the qubit state of $^9$Be$^+$ [dark ($|\uparrow\rangle_L$) or bright ($|\downarrow\rangle_L$)] is detected with up to 98 % fidelity by counting the fluorescence photons it scatters from the Doppler cooling laser within 200 µs (4). A threshold value discriminates between both states. The sequence is carried out multiple times (~100) within a few seconds with a fixed set of parameters to average the quantum projection noise and evaluate a mean excitation probability. To resolve linewidths approaching the natural linewidth of 17 Hz, a narrow-linewidth clock laser is required. Our home-built laser system is composed of a commercial extended cavity diode laser (ECDL) at 882 nm, prestabilised with a high locking bandwidth of 4 MHz to a passive external reference cavity using the ECDL pump current and grating piezo as actuators for the feedback. Thereby, we suppress laser high-frequency noise and obtain an instantaneous linewidth of about 2 kHz, limited by the relatively low cavity finesse (of about 1000) and lack of vibration-insensitive design. To suppress the residual noise, the laser is then further stabilised by phase-locking it to an ultra-stable laser operating at a wavelength of 1.5 µm. The latter is itself stabilised to a cryogenic cavity made from crystalline silicon (referred to as Si2)[37]. It achieves at averaging times between 1 and 50 seconds a fractional frequency instability at the thermal noise limit of the cavity of $4 \times 10^{-17}$. Using a femtosecond optical frequency comb as a transfer oscillator, we generate a virtual beatnote between the two lasers[38]. By demodulating it, we register their relative frequency and phase fluctuations, which are dominated by the significantly higher noise level of the 882 nm prestabilisation cavity. The demodulated beatnote is used to generate a feedback signal for phase-locking the two lasers, which is applied to an acousto-optic modulator between the ECDL and the prestabilisation cavity. The significantly lower bandwidth of the second locking stage ensures that the two loops do not compete with one another, but drifts and noise on the prestabilisation cavity out to the kHz-level are suppressed at the ECDL output, from which the spectroscopy light is derived. This suppresses the residual noise of the 882 nm clock laser, narrows its linewidth, and reduces the daily drift to a ~10 Hz level, dictated by Si2. The laser is frequency doubled to 441 nm in an external enhancement cavity containing a periodically-poled potassium titanyl phosphate (PPKTP) crystal. Active power stabilisation on a pulse-by-pulse basis at the ion trap was implemented for the clock, Doppler cooling, and Raman lasers to achieve stable system parameters such as Rabi frequencies and ac Stark shifts.

## Coherent laser spectroscopy of $^{40}$Ar$^{13+}$

Applying this technique, we carried out the first coherent laser spectroscopy of an HCI. Fig. 3a shows the excitation profile of the $m_{1/2} = -1/2$ to $m_{3/2} = -3/2$ Zeeman component of the $^{40}$Ar$^{13+}$ $^2$P$_{1/2}$-$^2$P$_{3/2}$ fine-structure transition. The blue curve shows a fit to the line by a Rabi line shape as expected for the top-hat laser pulse of 12 ms duration. This pulse length results in a 65 Hz FWHM Fourier-limited linewidth. We do not observe any additional line broadenings on this first-order Zeeman sensitive transition at this level, which confirms our previous measurements of a magnetic field stability of better than 1 nT[34] achieved via active stabilisation of the field in the vicinity of the vacuum chamber. Additionally, alternating external magnetic fields are shielded by the highly conductive cryogenic thermal shields made of high-purity copper with a low-pass corner frequency of ≤ 0.3 Hz and 30 to 40 dB suppression in the frequency range of 60 Hz to 1 kHz[34]. The maximum fringe contrast of about 0.4 at this probe duration was mostly limited by the excited-state lifetime, with contributions due to the approximately 90 % fidelity of the sideband operations on the two ions, imperfect state preparation and detection. Frequency scans with longer probe times can in principle resolve the natural linewidth, albeit at reduced excitation probability. Fig. 3b shows the on-resonance excitation probability as a function of the probe time for a higher intensity of the clock laser. Under continuous illumination, Rabi flopping between the two electronic states is observed (fitted by the red curve). The coherence decays with the known excited-state lifetime of 9.573(4) ms[39], indicated by the red-shaded exponential envelope of the fit. This measurement confirms coherence beyond this timescale for both the clock laser and the magnetic field. We also performed a direct measurement of the excited-state lifetime. For this, a carrier π-pulse (step (1) in Fig. 2) was applied with maximum laser intensity, which populated the $^{40}$Ar$^{13+}$ excited state in about 16 µs. After a variable wait time, the full transfer sequence was performed, and the remaining $^{40}$Ar$^{13+}$ excited-state fraction was mapped onto the $^9$Be$^+$ qubit state (see also Methods). During the wait time, a series of ground-state-cooling pulses on both axial motional modes was applied every millisecond to keep the two-ion crystal in the motional ground state in the presence of anomalous heating of 12 and 29 phonons per second for the OP and IP modes respectively. By incrementing the wait time in 1 ms steps, an axial mode temperature independent of the wait time is ensured. The observed exponential spontaneous decay of the excited state is shown in Fig. 3c and results in a lifetime of 9.97(26) ms. This is ~1.5 standard deviations longer than the more accurate experimental result of $9.573(4)_{\text{stat}} \binom{+12}{-5}_{\text{syst}}$ ms from an in-EBIT measurement[39] and advanced calculations of 9.538(2) ms[40] and 9.5354(20) ms[41]. The deviation from the previous measurement and calculations is consistent within the uncertainty. Further details are discussed in Methods.

## Measurement of the excited-state *g*-factor

A magnetic field of about 160 µT is applied at the location of the ions to define a quantisation axis and to deliberately split the Zeeman substates of $^9$Be$^+$ and $^{40}$Ar$^{13+}$ on a megahertz scale. With quantum logic-assisted HCI state preparation (see Fig. 2b and Extended Data Fig. 3), all six Zeeman components of the $^{40}$Ar$^{13+}$ $^2P_{1/2}$-$^2P_{3/2}$ transition can be coherently excited, as shown in Fig. 4a. The bottom x-axis represents the clock laser detuning from the degenerate line centre while the top x-axes represent the relative detunings from the centres of the individual Zeeman components. We can reconstruct the Zeeman shifts of the $^2P_{3/2}$ substates from this data (see Fig. 4b and c, as well as Methods), and derive the ratio of the *g*-factors of the excited and ground states from the measured frequencies. Within our current experimental precision of a few hertz over a splitting of several megahertz, we do not observe any quadratic contribution, which for instance could arise from electric field gradients coupling to the electric quadrupole moment of the $^2P_{3/2}$ state, or from a quadratic Zeeman shift. A quadratic term added to the fit function of Fig. 4b is consistent with zero. Recently, the ground-state *g*-factor $g_{1/2}$ of $^{40}$Ar$^{13+}$ was measured in the Penning trap experiment ALPHATRAP by the continuous Stern-Gerlach method to parts-per-billion accuracy[17]. Using this value, we obtain a weighted average of $g_{3/2}$ = 1.3322895(13)$_{stat}$(56)$_{syst}$ from three individual measurements (see Fig. 5). This is an improvement of more than two orders of magnitude over previous in-EBIT measurements[7], revealing for the first time the contributions to an HCI excited-state *g*-factor which arise from special relativity, interelectronic interactions, and QED. It also settles a discrepancy between previous theoretical values[42–46], confirming the configuration-interaction calculations of refs. [42,45,46] and very recent coupled-cluster calculations[47].

## Conclusions

We have cooled HCI to the ground state of motion in a linear Paul trap, making them the coldest HCI ever prepared in a laboratory. This enabled us to perform the first coherent, optical-clock-like laser spectroscopy of an electric dipole-forbidden optical transition in an HCI using quantum logic, at a level of precision that is eight orders of magnitude higher than the previous state-of-the-art. This proves the feasibility of hertz-level optical spectroscopy of HCI and opens up this large class of atomic systems to the tools of cutting-edge frequency metrology and quantum information processing.

The determination of the absolute frequency of the $^{40}$Ar$^{13+}$ fine-structure transition with a fractional uncertainty of $3\times10^{-15}$ and even higher levels of precision requires further evaluation of systematic shifts, such as the small time dilation shift from the residual motion of the ion[48] or the electric quadrupole shift[49], which is typically suppressed in HCI. By restricting measurements to the points of maximum frequency sensitivity of each line, frequency information can be obtained faster than when scanning the full line profiles as demonstrated here, further reducing the statistical uncertainty at a given averaging time[50]. At the same time, averaging over the Zeeman components on the second rather than minute timescale will suppress systematic uncertainties arising from drifting magnetic fields[51].

The presented techniques are not limited to our proof-of-principle HCI, $^{40}$Ar$^{13+}$, but can be applied more generally to forbidden transitions in other HCI. Several of the candidate species have properties that are even better suited for optical clock experiments, including much longer excited-state lifetimes and suppressed systematic shifts. Certain HCI are particularly sensitive to physics beyond the Standard Model such as a possible variation of the fine-structure constant[4], or to effects arising from fundamental interactions. A wide field of applications is the systematic study along isoelectronic sequences of relativistic effects in bound electronic systems and bound-state QED at ultra-high precision[5].

Furthermore, the techniques we have demonstrated here are not limited to the optical domain – this work also unlocks the new frontier of the vacuum ultraviolet and x-ray regimes for ultra-high precision spectroscopy, regions of the electromagnetic spectrum that are incompatible with neutral and singly-charged atoms due to unavoidable photoionisation. This will enable novel high-accuracy atomic clocks based on HCI and unrivalled tests of fundamental physics.

Note: During the revision of the manuscript a complementary work demonstrating incoherent laser spectroscopy of $^{40}$Ar$^{13+}$ in a Penning trap was published[52].

# Main text references

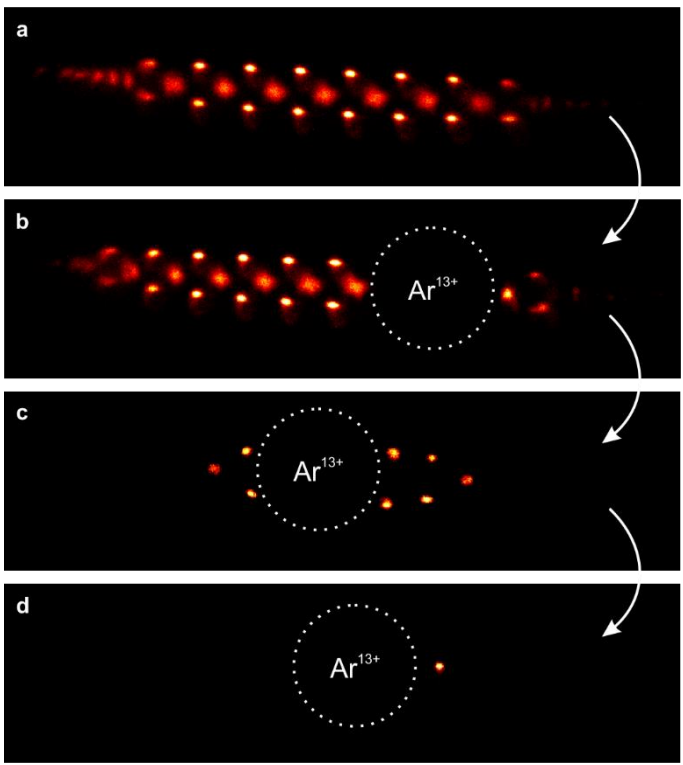

**Fig. 1 | Time sequence of highly charged ion recapture and two-ion crystal preparation. a**, A laser-cooled Coulomb crystal of 50 to 100 fluorescing $^9$Be$^+$ ions is confined in the Paul trap. **b**, A single Ar$^{13+}$ ion is injected along the crystal axis, sympathetically cooled, and finally co-crystallized. It appears as a large, dark void due to the repulsion of the $^9$Be$^+$ by the high charge state. **c**, Excess $^9$Be$^+$ ions are removed by modulating the Paul trap radio-frequency potential in the absence of laser cooling, resulting in heating and ion losses until, **d**, an Ar$^{13+}$-$^9$Be$^+$ two-ion crystal is prepared.

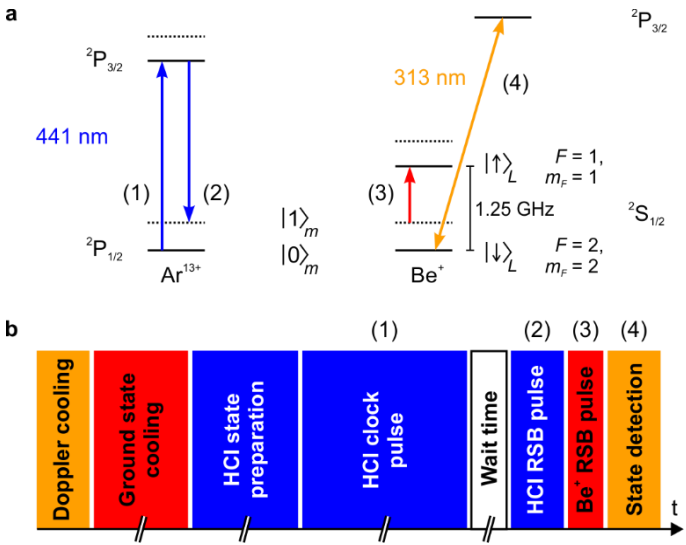

**Fig. 2 | Schematic illustration of the experimental cycle. a**, Level diagrams (not to scale) of boron-like Ar$^{13+}$ and $^9$Be$^+$. The motional Fock state of the crystal is denoted as $|n\rangle_m$. Solid and dotted black lines indicate the corresponding ground and first excited motional states, respectively. **b**, Experimental sequence. After Doppler cooling on the $^9$Be$^+$ $|\downarrow\rangle_L$-$^2P_{3/2}$ transition followed by ground state cooling to $|0\rangle_m$ by stimulated Raman transitions, the internal state of Ar$^{13+}$ is prepared (see also Fig. 4c and Extended Data Fig. 3). A clock laser pulse addressing the carrier of the Ar$^{13+}$ transition is then applied (1). After an optional wait time for lifetime measurements, a clock laser π-pulse on the Ar$^{13+}$ red sideband maps the Ar$^{13+}$ electronic state onto the common motional state (2). A red sideband π-pulse on the $^9$Be$^+$ hyperfine transition $|\downarrow\rangle_L$-$|\uparrow\rangle_L$ maps it onto the $^9$Be$^+$ electronic state (3). Finally, this state is detected using the Doppler cooling laser (4).

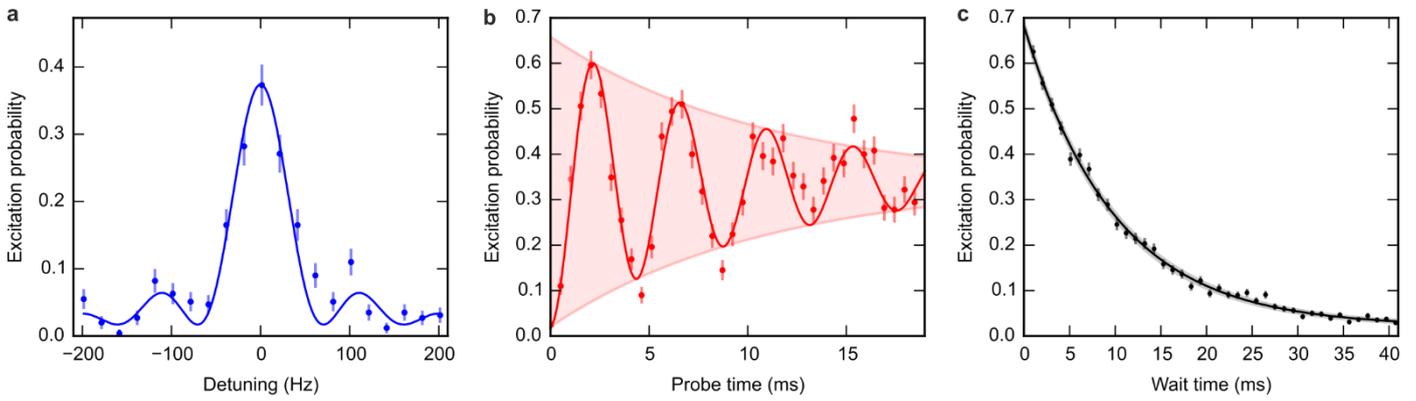

**Fig. 3 | Rabi spectroscopy and excited-state lifetime measurement. a**, Clock laser frequency scan across Zeeman component 1 (see Fig. 4c) of the $^{40}$Ar$^{13+}$ fine-structure transition. The fixed probe time of 12 ms is longer than the excited-state lifetime of 9.6 ms. The line is fitted by a Rabi line shape (blue curve), reaching a Fourier-limited FWHM of about 65 Hz. **b**, On-resonance coherent excitation of this transition. The coherent state Rabi flopping signal (fitted by the red curve, representing a damped sine with offset) exhibits a 2.2 ms π-time and decays exponentially with the lifetime of the excited state (red-shaded envelope). The error bars for **a** and **b** are given by the quantum projection noise of 255 measurements per data point. **c**, Excited-state lifetime measurement. Quantum logic sequences (see text) are carried out as a function of the wait time between carrier and red sideband clock laser pulses. During the wait time, the excited state can decay spontaneously. From a three-parameter maximum-likelihood estimation, we obtain a lifetime of 9.97(26) ms, limited by the quantum projection noise of 1100 measurements per data point (error bars). The black curve and grey-shaded area show the estimated exponential decay with the corresponding uncertainty band.

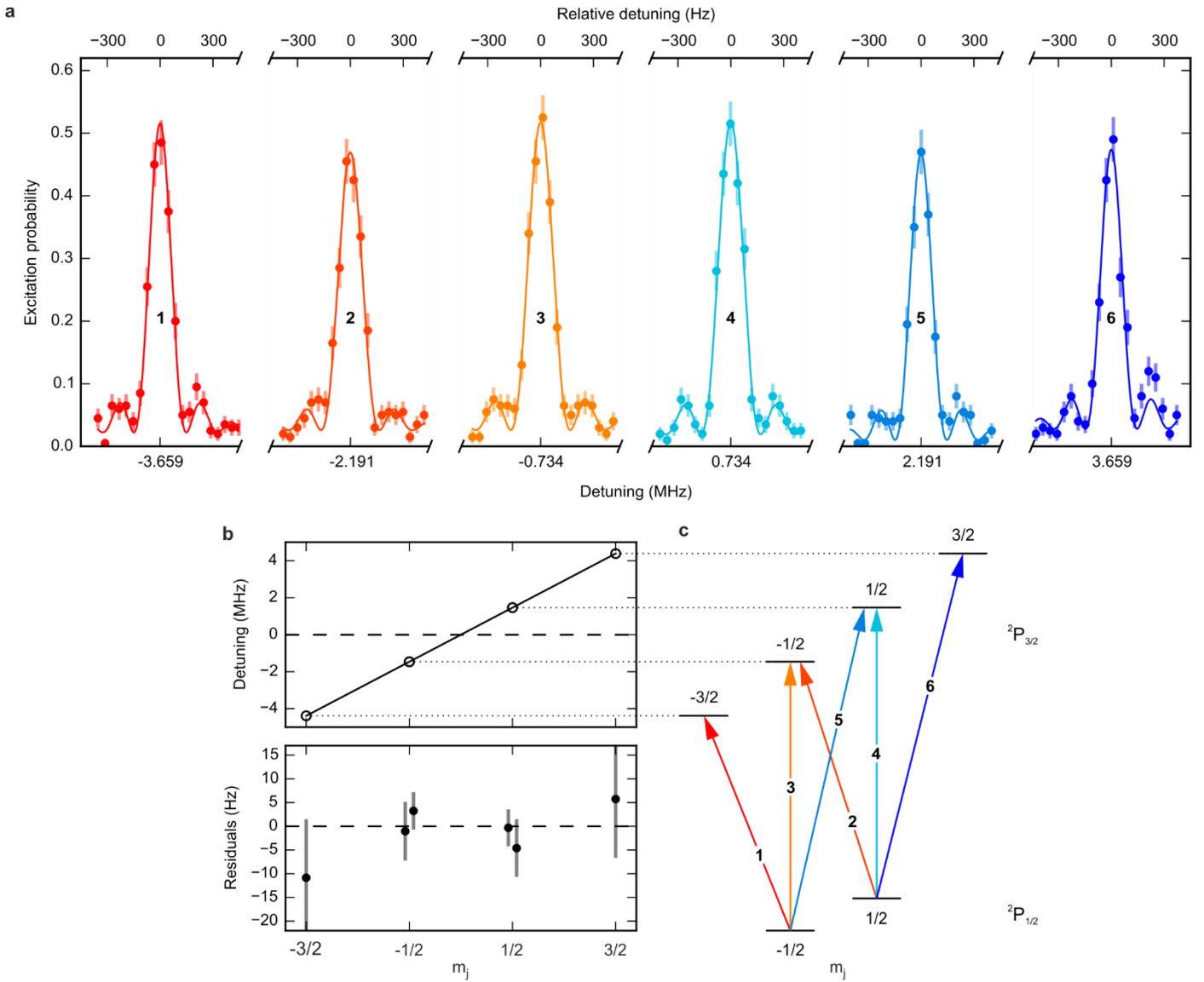

**Fig. 4 | Zeeman structure of the $^{40}$Ar$^{13+}$ $^2P_{1/2}$–$^2P_{3/2}$ fine-structure transition. a**, Excitation probability as a function of the clock laser detuning from the degenerate line centre, showing the six individual Zeeman components. Comparable laser-ion couplings (π-times between 5 and 6 ms) were chosen for each component, corresponding to Fourier-limited linewidths of around 150 Hz. The solid curves are Rabi line shape fit functions to determine the centre frequencies. Error bars represent the quantum projection noise of 200 repetitions. The varying excitation probabilities of the six components are caused by slightly different state preparation efficiencies. **b**, Reconstructed Zeeman shifts of the $^2P_{3/2}$ substates (upper panel) and their residuals (lower panel) with respect to a linear fit (solid black line). Note the different y-scales. For each of the substates $m_j = \mp 1/2$, two data points were obtained from transitions 2, 3 and 4, 5, respectively. A magnetic field instability of about 0.5 nT contributes to the standard uncertainties of the line centres, which become larger for the outer components. The slope of the linear fit is proportional to the ratio of the g-factors of the excited and ground states. **c**, Level diagram of the $^2P_{1/2}$ and $^2P_{3/2}$ Zeeman substates and corresponding Zeeman components of the fine-structure transition.

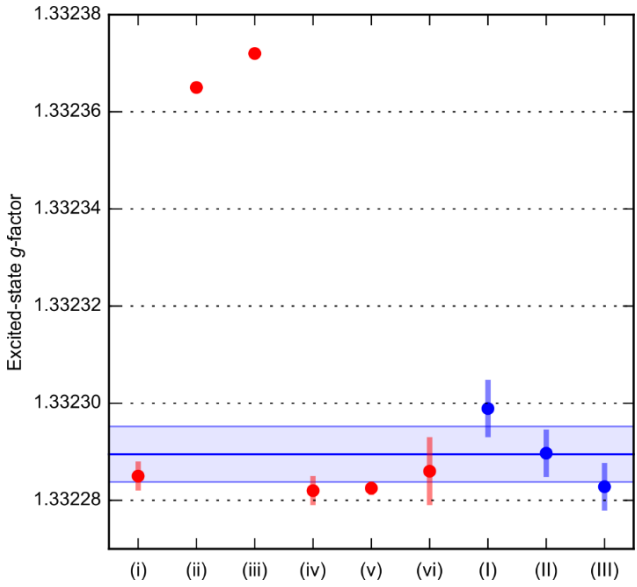

**Fig. 5 | Comparison of the calculated (red) and measured (blue) excited-state *g*-factor**. (i) – Glazov *et al., Phys. Scr.* (2013)[42], (ii) – Verdebout *et al., At. Data Nucl. Data Tables* (2014)[43], (iii) – Marques *et al., Phys. Rev. A* (2016)[44], (iv) – Shchepetnov *et al., J. Phys. Conf. Ser.* (2015)[46], (v) – Agababaev *et al., X-Ray Spectrom.* (2019)[45], (vi) – Maison *et al., Phys. Rev. A* (2019)[47]. The error bars of (iii) and (v) are smaller than the data points. (ii) does not provide any uncertainty. I, II, and III are the evaluated *g*-factors from the three data sets produced in this work with their standard uncertainty. The solid blue line displays the weighted average with the 1σ uncertainty band, with the largest contribution coming from the systematic uncertainty. See Methods for details.

## METHODS

**Highly charged ion production, transfer and recapture.** We show a top view of the laboratory setup in Extended Data Fig. 1, and a simplified schematic of the potential landscape in Extended Data Fig. 2a. Highly charged ions (HCI) are produced by electron impact ionisation and stored by PTB-EBIT, a Heidelberg-type compact electron beam ion trap (EBIT)[32]. After extraction of ions in bunches, a beamline with multiple electrostatic elements is used to guide them towards the Paul trap and to manipulate their kinetic energy distribution. Five segmented einzel lenses[53] and an electrostatic double-focusing 90° deflector[54] are employed for focusing and steering. A pair of pulsed drift tubes (following the approach described in ref. [33]) is used for deceleration and pre-cooling, reducing the phase-space volume of the bunches. Downstream, two microchannel plate (MCP) detectors can be moved into the ion beam in front of and behind the Paul trap to optimise ion yield and beam transmission. The first one also features a retarding field analyser used for determining the mean kinetic energy and the energy spread of the ion bunches. While the present method of HCI production, transfer, and recapture combining EBIT, beamline, and Paul trap can handle a large variety of elements and charge states, the following section refers specifically to the present case of optimised $^{40}\text{Ar}^{13+}$ recapture.

Highly charged argon ions are produced in a distribution of charge states by means of a 13 mA, 1 keV electron beam in an approximately 50-V-deep axial trapping potential of the EBIT. In each cycle, the central trap electrode is rapidly switched from about 450 V (for the aforementioned 50-V-deep trap) to a repulsive extraction potential of 700 V (200 V higher than the outer trap electrode potential) for a period of 200 ns at a rate of 4 Hz to eject the ions. The kinetic energy relative to the ground potential of the beamline (0 V) and thus the velocity of extracted ions depend on the total extraction potential of 700 V as well as on the ionic charge $q$, allowing a separation of the different charge states by their different time-of-flights (Extended Data Fig. 2a and b). $^{40}\text{Ar}^{13+}$ is selected with the help of an electrode of the third segmented einzel lens just behind the 90° deflector. This electrode acts as a gate by rapidly switching it to a passing voltage at the $^{40}\text{Ar}^{13+}$ arrival time and back to a deflecting voltage after the ion passage. Thus, the trajectories of all other charge states are deflected away from the Paul trap. We measured a mean kinetic energy of 694 $q$V with respect to ground for the fast $^{40}\text{Ar}^{13+}$ bunches by means of the retarding field analyser (Extended Data Fig. 2e). An associated axial energy spread of 32 $q$V was also determined. To decrease the mean kinetic energy and its spread to values more amenable for trapping and efficient cooling in the cryogenic Paul trap, we perform an electrodynamic deceleration step with the pair of pulsed drift tubes. By biasing them to approximately 510 V and 590 V respectively before the extraction, a linear axial potential gradient is generated on the beamline axis between the two electrodes. When the ion bunch arrives at that position about 9.7 µs after ion ejection, it is thus exposed to a mean potential of 550 V. Then, both drift tube potentials are rapidly grounded using a fast high-voltage switch. This slows down the ion bunch to a kinetic energy of 146 $q$V and reduces the axial energy spread to 13 $q$V (Extended Data Fig. 2f). The deceleration step also significantly shortens the length of the ion bunches from about 5.2 cm to about 1.7 cm full-width-at-half-maximum (FWHM), while their temporal width is only slightly reduced (Extended Data Fig. 2c and d). After passing through a final einzel lens and an unbiased mirror tube, the $^{40}\text{Ar}^{13+}$ ions enter the Paul trap.

Its voltages are commonly biased to 138 V to accomplish the final electrostatic deceleration step. This brings the ions to a residual kinetic energy of about 5 to 10 $q$V.

The Paul trap is formed by a radially confining radio-frequency (rf) potential and an axially confining dc potential. Once inside the trap, the HCI repeatedly pass through a cigar-shaped Coulomb crystal of about 50 to 100 $^9$Be$^+$ ions that has been previously loaded into the Paul trap by means of laser ablation combined with photoionisation[34] (see also Fig. 1). This proceeds as follows. During injection into the Paul trap, owing to their relatively high kinetic energy, most HCI can overcome the weak axially confining potential of 300 meV (above the biased ground of 138 V) applied to the electrostatic endcap at the entrance of the trap. After passing through the $^9$Be$^+$ Coulomb crystal for the first time, the $^{40}$Ar$^{13+}$ ions are reflected by the opposite electrostatic endcap potential of about 12 V (above the biased ground of 138 V). In the meantime (17.1 µs after initial ion extraction from the EBIT), the mirror tube at the entrance of the Paul trap is rapidly switched up to a confining axial electrostatic potential to complete the capture of $^{40}$Ar$^{13+}$. Then the trap remains closed for 1.9 s, during which the HCI can dissipate their residual kinetic energy by repeated interactions with the laser-cooled $^9$Be$^+$ ions. If these steps are successful for an $^{40}$Ar$^{13+}$ ion, it joins the $^9$Be$^+$ Coulomb crystal (Fig. 1). Otherwise, the mirror-tube potential is lowered again to let the next HCI bunch enter the Paul trap. This whole recapture process is rather efficient and succeeds on average in less than 30 s.

**Excited-state lifetime measurement.** The data for the lifetime measurement include 440 measurements, each of which with 100 experimental realisations, adding up to a total measurement time of about two hours. Eleven measurements were averaged for every single waiting time, with the error bars in Fig. 3c indicating the quantum projection noise of 1100 experimental implementations. In order to cancel the effects of parameter drifts on the observed signal, the waiting time was scanned in a pseudo-random sequence. Drifts of the atomic resonance frequencies could lead to systematic variations in the detected excitation probabilities. The shortest achievable π-times of 16 µs for the initial HCI excitation and 225 µs for the HCI sideband transition lead to an interaction-broadening of the respective lines of 62 kHz and 4.4 kHz, respectively. Our typical short-term magnetic field fluctuations lead to line shifts on the < 10 Hz level, and thus affect the measured excited-state population on the order of $10^{-4}$. The axial trap frequency has fluctuations below 100 Hz over the course of a day. The distribution of data points for a given wait time is consistent with the expected quantum projection noise, thereby ruling out systematic drifts at the level of the statistical uncertainty.

The clock laser pulses are generated by the first diffraction order of an acousto-optic modulator (AOM). Despite the typical 100 dB level of extinction of the rf drive power provided by an active rf switch, the optical extinction ratio does not reach this level due to scattered light within the AOM crystal. However, this leaked light is unshifted by the AOM and therefore detuned from the ion resonance by the rf drive frequency of about 200 MHz, or $10^7$ natural linewidths. This alone reduces the de-excitation probability by approximately 14 orders of magnitude.

Spontaneous decay of the HCI on the red sideband is suppressed by the square of the Lamb-Dicke parameter, $\eta^2 \approx 0.01$. However, residual decay on this sideband leads to heating of the motional mode and may thus appear as spurious excitation in the quantum logic detection. A few sideband cooling pulses applied just before the quantum logic transfer pulse suppress this effect by returning the crystal to its ground state. Off-resonant depumping of the excited state of $^{40}\text{Ar}^{13+}$ by the $^9\text{Be}^+$ lasers is negligible due to the narrow natural linewidth and the large detuning. Collisional de-shelving as discussed in refs. [55,56] is absent in this experiment due to the extreme high vacuum. Furthermore, collisions of an HCI with a neutral particle likely lead to charge exchange and total, but inconsequential ion loss.

**g-factor evaluation.** The $^{40}\text{Ar}^{13+}$ $^2P_{3/2}$ excited-state g-factor, denoted as $g_{3/2}$, is determined by a linear fit of the Zeeman substate energy shifts. We use the well-known ground-state g-factor of the clock transition from a recent high-accuracy measurement[17] to operate a co-magnetometer and measure the magnetic field by an appropriate combination of the Zeeman components.

The energy shifts $\Delta E_{3/2,m_{3/2,i}} = m_{3/2,i}\, g_{3/2}\mu_B B$ of the Zeeman substates of the excited $^2P_{3/2}$ state due to an external magnetic field $B$ are obtained from the measured Zeeman shifts $f_i$ (in units of frequency) of the six Zeeman components ($i$ ranging from 1 to 6 according to Fig. 4c) and analogously the shifts $\Delta E_{1/2,m_{1/2,i}}$ of the $^2P_{1/2}$ Zeeman substates. $h$ and $\mu_B$ are the Planck constant and the Bohr magneton, respectively. The shifts are referenced with respect to the degenerate line/level centres. One then obtains:

$$\frac{\Delta E_{3/2,m_{3/2,i}}}{h} = f_i + \frac{\Delta E_{1/2,m_{1/2,i}}}{h} \qquad (1)$$

$$\Leftrightarrow m_{3/2,i}\, g_{3/2} \frac{\mu_B B}{h} = f_i + m_{1/2,i}\, g_{1/2} \frac{\mu_B B}{h}. \tag{2}$$

$B$ is eliminated from the above equation by using the four inner Zeeman components 2-5, which are less sensitive to magnetic field fluctuations than the two outer ones. Components $f_2, f_3$ share the common excited state $m_{3/2} = -1/2$ (see Fig. 4c), and therefore their difference directly yields the ground-state Zeeman splitting without relying on the excited-state $g$-factor. With the known ground-state $g$-factor, $g_{1/2}$ from the work of Arapoglou et al.[17], we obtain the magnetic field

$$B_1 = \frac{h(f_3 - f_2)}{g_{1/2}\, \mu_B}. \tag{3}$$

Similarly, components $f_4, f_5$ share the excited state with $m_{3/2} = +1/2$ and we acquire a second measurement of

$$B_2 = \frac{h(f_5 - f_4)}{g_{1/2}\, \mu_B}. \tag{4}$$

Introducing $U = f_5 - f_4 + f_3 - f_2$ for simplicity, we average the magnetic field $B$ from these two relations to reduce the uncertainty:

$$B = \frac{B_1 + B_2}{2} = \frac{h\, U}{2\, g_{1/2}\, \mu_B}. \tag{5}$$

This expression is inserted into Eq. (2) to obtain:

$$y_i(m_{3/2,i}) := \frac{g_{3/2}\, U}{2\, g_{1/2}}\, m_{3/2,i} = f_i + m_{1/2,i}\, \frac{U}{2}. \tag{6}$$

On the right-hand side of the equal sign of this equation, the measured shifts of the excited Zeeman substates are given and fulfill a linear relation in $m_{3/2,i}$ (left-hand side of the equation). A linear fit (see black line in Fig. 4b) of the form

$$y_i(m_{3/2,i}) = a \cdot m_{3/2,i} + b \tag{7}$$

with offset $b$ to account for a global frequency offset in the measured $f_i$ allows us to determine the excited-state $g$-factor $g_{3/2}$ from the slope $a$:

$$g_{3/2} = \frac{2\, g_{1/2}\, a}{U}. \tag{8}$$

The uncertainties $\sigma_{y_i}$ of the excited Zeeman substates are obtained from the right-hand side of Eq. (6) by expressing $U$ again through $U = f_5 - f_4 + f_3 - f_2$, followed by standard uncertainty propagation with the independently measured $f_i$:

$$\sigma_{y_i} = \sqrt{\sum_j \left(\frac{\partial y_i}{\partial f_j} \sigma_{f_j}\right)^2} \qquad (9)$$

The uncertainties $\sigma_{f_i}$ of the Zeeman components depend on the statistical uncertainty of the line centre from the fit, $\sigma_{f_{i,fit}}$ (fitting the lines by Rabi line shapes), and the relative systematic magnetic field uncertainty $\sigma_B/B$. The latter is time-dependent and is estimated from the observed magnetic field stability previously measured by using the $^9$Be$^+$ qubit transition frequency (see reference [28] for details) to be $4.1\times10^{-6}$ (measurement 1) and $3.2\times10^{-6}$ (measurements 2 and 3) on relevant time scales. Accordingly, one has:

$$\sigma_{f_i} = \sqrt{\sigma_{f_{i,fit}}{}^2 + \left(\frac{\sigma_B}{B} f_i\right)^2} \qquad (10)$$

The linear fit shown in Fig. 4b is weighted with the $\sigma_{y_i}$ uncertainties, which are displayed in the lower panel. For completeness, we state the fit offsets of $b = -17(3)$ Hz, $-45(2)$ Hz, and $-30(2)$ Hz for the three sets of measurements. The reduced $\chi^2$ of the linear fits are 1.45, 0.57, and 0.21.

To estimate the uncertainty of $g_{3/2}$, we replace $a$ and $U$ in Eq. (8) by their analytical expressions. $a = \overline{(m_{3/2,i} \cdot y_i)} / \overline{m_{3/2,i}{}^2}$, obtained from the closed-form solution of a linear fit. The $y_i$ are given by the right-hand side of Eq. (6), and $U = f_5 - f_4 + f_3 - f_2$. We can neglect the parts-per-billion uncertainty of the experimental result $g_{1/2} = 0.66364845532(93)$ from the very recent Penning trap measurement[17], since it is more than three orders of magnitude smaller than our experimental uncertainties in the 25000 times weaker magnetic field. Finally, the only uncertainties are introduced by the independently measured $f_i$. Thus, the uncertainty $\sigma_{g_{3/2}}$ is obtained from the common formula of uncertainty propagation

$$\sigma_{g_{3/2}} = \sqrt{\sum_i \left(\frac{\partial g_{3/2}}{\partial f_i} \cdot \sigma_{f_i}\right)^2}. \qquad (11)$$

The evaluated $g_{3/2}$ values are 1.3322989(19)$_{stat}$(56)$_{syst}$, 1.3322897(23)$_{stat}$(43)$_{syst}$, and 1.3322828(24)$_{stat}$(43)$_{syst}$ from the three measurement sets obtained on two different days, where we have stated the statistical and systematic uncertainties separately. The results are shown in Fig. 5 together with recent calculations. The uncertainties of the individual measurements are the root of the sum of the squared statistical and systematic uncertainties. The measurements agree within their uncertainties, where the largest deviation between measurement 1 and the weighted average is 1.6 standard deviations. We obtain the weighted average $g_{3/2}$ = 1.3322895(13)$_{stat}$(56)$_{syst}$, where we have combined the statistical uncertainties and stated the largest systematic uncertainty of the individual measurements as a conservative estimate for the systematic uncertainty of the average.

## Methods references

**Acknowledgements** We acknowledge Ioanna Arapoglou, Hendrik Bekker, Sven Bernitt, Klaus Blaum, Alexander Egl, Stephan Hannig, Steffen Kühn, Thomas Legero, Robert Müller, Janko Nauta, Julian Stark, Uwe Sterr, Sven Sturm, Andrey Surzhykov, and the electronics and mechanical workshops of QUEST, and of the Max-Planck-Institut für Kernphysik, as well as divisions 4 and 5.5 of the PTB for support and helpful discussions. The project was supported by the Physikalisch-Technische Bundesanstalt, the Max-Planck Society, the Max-Planck–Riken–PTB–Center for Time, Constants and Fundamental Symmetries, and the Deutsche Forschungsgemeinschaft (DFG, German Research Foundation) through SCHM2678/5-1, the collaborative research centres SFB 1225 ISOQUANT and SFB 1227 DQ-*mat*, and Germany's Excellence Strategy – EXC-2123/1 QuantumFrontiers. This project also received funding from the European Metrology Programme for Innovation and Research (EMPIR) co-financed by the Participating States and from the European Union's Horizon 2020 research and innovation programme (Project No. 17FUN07 CC4C). S.A.K acknowledges financial support from the Alexander von Humboldt Foundation.

**Author Contributions** P.M., T.L., S.A.K., E.B., L.S., M.S., J.R.C.L.-U., and P.O.S. developed the experimental setup. P.M., T.L., S.A.K., and L.J.S. carried out the experiments. P.M. and T.L. analyzed the data. J.R.C.L.-U. and P.O.S. conceived and supervised the study. P.M. and P.O.S. wrote the initial manuscript with contributions from T.L., S.A.K., and J.R.C.L.-U.. All authors discussed the results and reviewed the manuscript.

**Author Information** Reprints and permissions information is available at www.nature.com/reprints. The authors declare no competing interests. Correspondence should be addressed to P.M. (Peter.Micke@quantummetrology.de) and P.O.S. (Piet.Schmidt@quantummetrology.de).

**Data Availability** The datasets generated and analysed during this study are available from the corresponding author upon reasonable request.

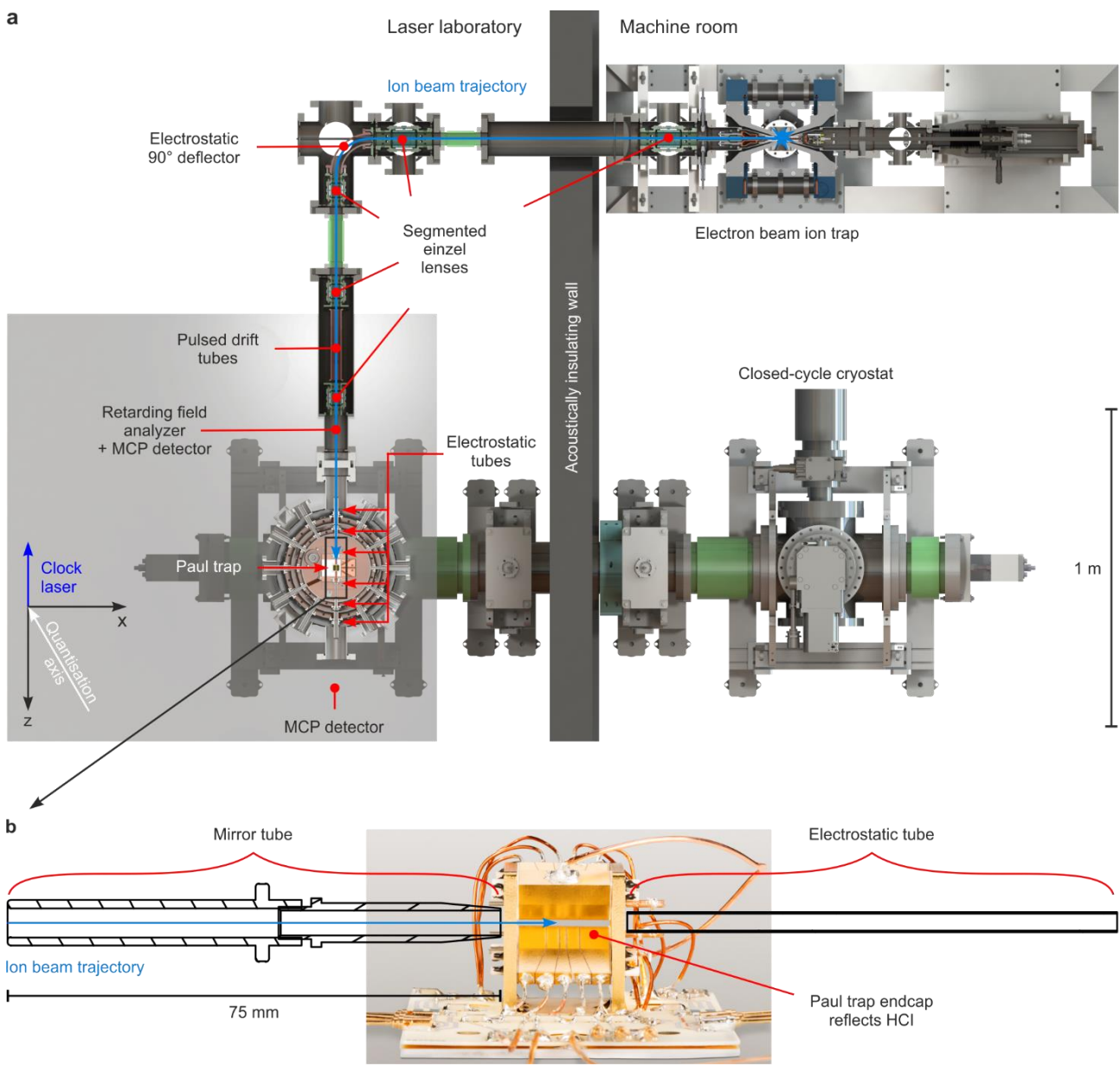

**Extended Data Fig. 1 | Experimental setup. a**, Top view. The apparatus extends over two rooms separated by an acoustically insulating wall. Inside the 'machine room' on the right-hand side, HCI are produced in an electron beam ion trap[32] and extracted as ion bunches along the ion beam trajectory (blue line) through a deceleration beamline. At the laser laboratory (left side) they are axially injected into a cryogenic linear Paul trap[34], which is mounted on a pneumatically floating optical table (grey-shaded). The Paul trap is refrigerated by a vibrationally decoupled pulse tube cryocooler[35] located in the machine room. The beamline is composed of several ion-optical elements: five segmented einzel lenses as well as an electrostatic 90° deflector for guiding and focusing the ions, a pair of pulsed drift tubes for deceleration, and six cylindrical electrodes arranged in line in front of and behind the Paul trap. Charge-state separation is accomplished by the different time-of-flights through the beamline. One electrode of the third segmented einzel lens is used as a gate to select the desired charge state. A microchannel plate (MCP) detector in front of the Paul trap includes two fine stainless-steel meshes to apply a well-defined retarding field and allows a measurement of the kinetic energy distribution of the ion bunches (see also Extended Data Fig. 2e and f). A second MCP detector behind the Paul trap is used to optimize the ion beam transmission through the Paul trap. **b**, Magnified side view of the cryogenic Paul trap region. The trap (photograph) is shown with the two adjacent electrostatic tubes. The left one (mirror tube) at the entrance of the Paul trap is used to capture the HCI by rapidly switching it to a confining potential once the HCI have passed it.

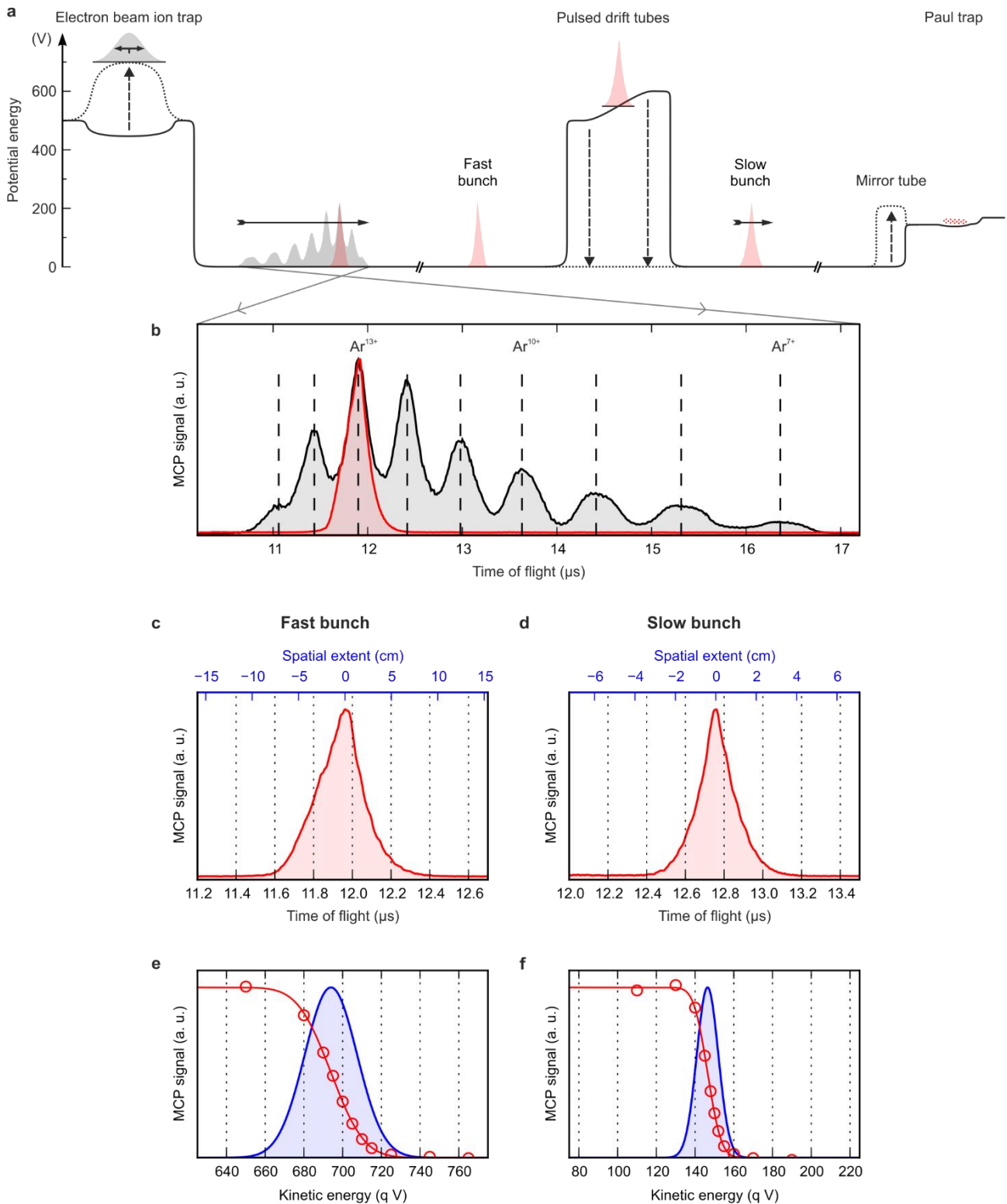

**Extended Data Fig. 2 | Highly charged ion extraction and transfer. a**, Simplified illustration of the electrostatic potential for the $^{40}$Ar$^{13+}$ transfer from the electron beam ion trap (EBIT) to the Paul trap. The entire ion inventory stored in the EBIT, with its charge-state distribution displayed as gray-shaded, is ejected by switching the axial trap to a repulsive potential. The charge states separate due to their distinct initial kinetic energy. $^{40}$Ar$^{13+}$ ions (red) are selected by an electrode used as a gate (not shown). The fast $^{40}$Ar$^{13+}$ bunch is then slowed down upon entering the pulsed drift tubes. Having arrived there at the centre of a linear potential gradient, the electrode potentials are rapidly switched to ground, and a slower $^{40}$Ar$^{13+}$ bunch leaves the pulsed drift tubes. At the Paul trap, the ions are further

decelerated by an electrostatic potential and enter the trapping region with a reduced residual kinetic energy of 5 to 10 $q$V. They then pass a Coulomb crystal of $^9$Be$^+$ ions and are reflected by an electrostatic endcap electrode biased to a potential of about 12 V above the biased common ground. Meanwhile, an electrostatic mirror tube in front of the Paul trap has been switched up to a confining potential at which $^{40}$Ar$^{13+}$ is unable to escape the Paul trap. This causes an oscillatory motion along the trap axis. Through the repeated interactions with the laser-cooled $^9$Be$^+$ ions, $^{40}$Ar$^{13+}$ dissipates its residual kinetic energy and joins the Coulomb crystal. **b**, Normalised ion yield as a function of the time-of-flight after ion ejection from the EBIT, measured by the first microchannel plate (MCP) detector in front of the Paul trap. The black curve shows the entire charge state distribution, with Ar charge states from +7 through +15. Using the gate electrode, $^{40}$Ar$^{13+}$ is chosen for passage, as shown by the red curve. **c**, Normalised fast $^{40}$Ar$^{13+}$ bunch as a function of time (FWHM of about 250 ns) and position along the beamline axis (averaged over 16 shots) and, **d**, for a normalised slow $^{40}$Ar$^{13+}$ bunch (FWHM of about 185 ns). **e**, Normalised kinetic energy distribution of the fast $^{40}$Ar$^{13+}$ bunch along the beamline axis and, **f**, of the slow bunch after deceleration and phase-space cooling using the pulsed drift tubes. The red circles show the integrated ion yield of an averaged $^{40}$Ar$^{13+}$ bunch (16 shots) for a given retardation potential, measured by the retarding-field analyser. A Gaussian error function (red line) was fitted to the data and differentiated to obtain the Gaussian energy distribution (blue line) to show the mean kinetic energy and longitudinal energy spread.

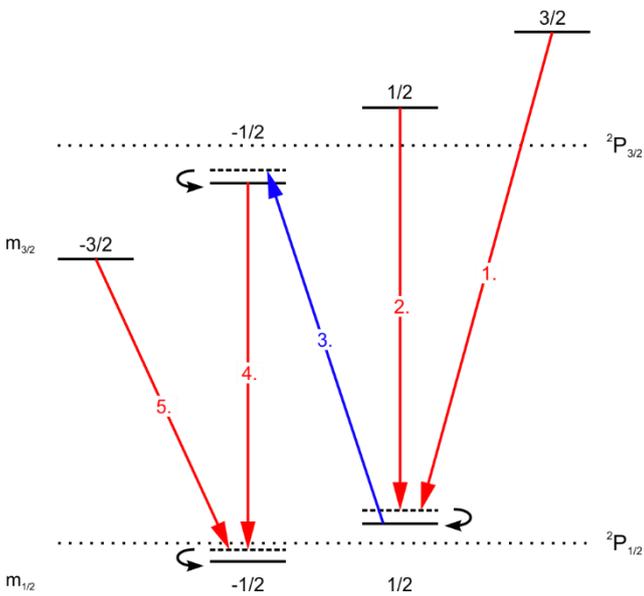

**Extended Data Fig. 3 | Quantum logic-assisted internal state preparation of Ar$^{13+}$.** The $m_{1/2} = -1/2$ state of the $^2$P$_{1/2}$ level is deterministically populated by a series of five clock laser sideband π-pulses (1. through 5.), which excite the two-ion crystal from the motional ground state $|0\rangle_m$ (solid lines) into the excited state $|1\rangle_m$ (dashed lines). By means of Raman sideband cooling pulses acting on the $^9$Be$^+$ ion, the crystal is returned to the motional ground state after each

transfer pulse. This ensures unidirectional optical pumping[9]. To increase the state-preparation efficiency, this sequence is repeated four times. The other Zeeman ground state ($P_{1/2}$, $m_{1/2} = +1/2$) is prepared in an analogous manner.